# Pressure-induced enhancement of superconductivity in a non-centrosymmetric compound LaPtGe

Sathiskumar Mariappan[1,2], Dilip Bhoi[1], Boby Joseph[3], R. P. Singh[4], Govindaraj Lingannan[2], Velaga Srihari[5], Arumugam Sonachalam[2*] and Yoshiya Uwatoko[1*]

[1]Institute for Solid State Physics, The University of Tokyo, 5-1-5 Kashiwanoha, Kashiwa, Chiba, 277-8581, Japan.
[2]Centre for High Pressure Research, School of Physics, Bharathidasan University, Tiruchirappalli-620024, India.
[3]Elettra-Sincrotrone Trieste S.C. p. A., S.S. 14, Km 163.5 in Area Science Park, 34149 Basovizza, Italy.
[4]Indian Institute of Science Education and Research Bhopal, Madhya Pradesh, 462066, India.
[5]High Pressure and Synchrotron Radiation Physics Division, Bhabha Atomic Research Centre, Mumbai, India.

***Corresponding authors:** sarumugam1963@yahoo.com, uwatoko@issp.u-tokyo.ac.jp

**Abstract: -**

We report a pressure-induced enhancement of the superconducting transition temperature ($T_c$) in a non-centrosymmetric (NCS) compound, LaPtGe. With pressure, till 3 GPa, we observed a modest enhancement of the $T_c$ with a rate of 0.071 K/GPa. However, above this pressure, the rate showed a ~2.5 times increase, 0.183 K/GPa. We observed a $T_c$ of 3.94 K at 6 GPa, the highest pressure value used in our transport study. Synchrotron high-pressure x-ray powder diffraction (HP-XRPD) measurements do not reveal any structural phase transition in this system in this pressure range. However, it showed an apparent change of slope in the pressure dependence of lattice parameters above and below 3 GPa. Pressure dependence of the unit-cell volume also followed a distinct trend below and above 3 GPa, with the Birch-Murnaghan equation of state fit providing a bulk modulus ($B_0$) value of ~144 and ~162 GPa, respectively, for two pressure regions. Further, the magnetotransport measurement under pressure up to 2.45 GPa reveals the enhancement of the upper critical field ($H_{c2}(0)$) from 0.7 T (0 GPa) to 0.92 T (2.45 GPa). In addition, the upward curvature in $H_{c2}(T)$ becomes stronger with increasing pressure, suggesting a change of the underlying Fermi surface topology with pressure. The signature for the inducible $H_{c2}(T)$ with pressure, the distinct enhancement of $T_c$ around 3 GPa, and the noticeable change in lattice parameters around 3 GPa suggests the possibility of multi-gap superconductivity in LaPtGe similar to identical structure NCS compound, LaPtSi. The enhancement of $T_c$ by pressure can be correlated with the possible underlying lattice modulation by compression and the change in the density of states at the Fermi level. Also, the bare change in pressure-dependent activation energy $U_0$ up to 2.45 GPa calculated using the Arrhenius relation clearly shows that the shift in $T_c$ does not arise from grain boundaries.

# Introduction: -

Non-centrosymmetric (NCS) superconductors have attracted significant interest due to the presence of a non-uniform lattice potential and spin degeneracy, which results in antisymmetric spin-orbit coupling (ASOC) [1–3]. The occurrence of ASOC can cause the coexistence of spin-singlet and spin-triplet pairing in many NCS superconductors [4–11]. This mixed pairing scenario may be reflected in several physical properties, such as the nodal superconducting gap, large upper critical field at absolute zero ($H_{c2}(0)$) exceeding the Pauli upper critical field ($H_P$) and the time-reversal symmetry breaking (TRSB). The signature of this mixed pairing state is observed in both magnetic and non-magnetic NCS compounds. Among that, it is claimed that the magnetic fluctuations or strong correlation effect of *f*-electrons might be the reason for the mixed paring state of the magnetic NCS compounds, for example, $CeCu_2Si_2$ [12], $CePt_3Si$ [11], $CeIrSi_3$ [11], $CeRhSi_3$ [13,14], $CeCoGe_3$ [15], UIr [16,17], $CeCo_2$ [18], $CeRhIn_5$ [19], $CeCoIn_5$ [20], $CeNiC_2$ [21] and $CeIrGe_3$ [11]. Therefore, the non-magnetic NCS superconductors are promising materials for understanding the role of crystal inversion symmetry in the mixed pairing state.

In this context, a non-magnetic NCS LaPtGe ($T_c \sim 3.0$ K) is an exciting system to study due to its s-wave superconductivity with the absence of TRSB, which crystallizes in a body-centred tetragonal *α-ThSi$_2$* type structure with space group *I4$_1$md*, No.109 [22]. It is noted that a recent zero-field muon spin rotation (*ZF-μSR*) study shows the TRSB in LaPtGe [23]. However, the report [22] shows the preserved TRSB and the *ZF-μSR* spectra exhibit a slight but clear difference between 0.3 and 4.5 K, which does not entirely rule out the TRSB. Further, the calculations of electronic band structure illustrate a stronger ASOC in LaPtGe than the LaNiSi and LaPtSi [24]. Even though the signature-like multi-gap nodal superconductivity is not observed in LaPtGe, unlike LaPtSi [24] [23] and other similar family of compounds like LaNiSi [23] and CaPtAs [25] [26]. So, it is crucial to determine the essential role of the lack of inversion symmetry in the mixed pairing state or other unconventional properties of LaPtGe. Besides, recently, it has been predicted that LaNiSi, LaPtSi, and LaPtGe are Weyl nodal-line semimetals due to the presence of a non-symmorphic gliding plane [24]. Further, replacing of Pr atom in the Ce site of CeGaGe will enhance the $T_c$ [27]. Apart from the atom replacement, the external pressure is one of the clean and proven ways to control the superconducting transition temperature ($T_c$) of material by influencing the Fermi surface density of states (DOS) as well as electron-phonon coupling by subtle variation of crystal structural parameters. And it was observed in many NCS compounds; for example, the enhancement of superconductivity with pressure was observed in Ce(Ir, Rh)Si$_3$ [11,14], CeIrGe$_3$ [11], CeCoGe$_3$ [15], UIr [16], $Cd_2ReO_7$ [28], and $Rh_2Mo_3N$ [29,30].

Keeping these points in mind, we have investigated the superconducting and structural properties of LaPtGe [22] under pressure through high-pressure (HP) electrical transport measurement up to 6 GPa and HP-X-ray powder diffraction (XRPD) up to 7.8 GPa measurement. The enhancement of superconducting transition temperature ($T_c$), superconducting transition width ($\Delta T_c$) and decrease of residual resistivity ratio (RRR) under pressure (P) with notable change around 3 GPa from the temperature dependence of HP resistivity ($\rho(T)$). Further, the $\rho(T)$ under the external magnetic field shows an enhancement of the upper critical field at zero Kelvin $H_{c2}(0)$. The HP-XRPD, up to 7.82 GPa at room temperature, shows the decrease of lattice parameters and unit cell volume without any structural phase transition. From HP-XRPD, below and above ~3 GPa, two different pressure-dependent compression rates were observed, with a slight change in the bulk modulus ($B_0$) ~144 GPa and ~162 GPa, respectively. Interestingly, this peculiar pressure-dependent compression of a unit cell is relevant to the enhancement of pressure-dependent $T_c$ and $H_{c2}(0)$. Also, the field dependence of activation energy $U_0(H)$ is analyzed in the thermally activated flux flow (TAFF) region, confirming the transition of single to collective vortex states.

## Experimental Details: -

A polycrystalline sample of LaPtGe was synthesized from the high purity (99.99 %) La, Pt, and Ge using the arc melting method. The sample phase purity was confirmed by laboratory X-ray diffraction studies [22]. The electrical resistivity measurements were done using the standard linear four-probe method, and the contacting leads were made by gold wire ($\phi$ 20 mm) and high-purity silver epoxy. Further, the $\rho(T)$ at various fixed magnetic fields and field dependence of resistivity ($\rho(H)$) at different temperature points were measured both at ambient and high pressures up to 2.45 GPa using PPMS. The HP up to 2.45 GPa was attained using the clamp-type highbred piston-cylinder pressure cell made of Be-Cu and Ni-Cr-Al and the Daphne 7474 oil (Idemitsu Kosan Co. Ltd.) as a pressure transmitting medium (PTM) to ensure the hydrostaticity [31]. The actual pressure inside the pressure cell was calculated from the $T_c$ of Pb, which was mounted along with the sample [32]. The self-clamp palm-type cubic anvil cell (P-CAC) was utilized for high-pressure resistivity measurements up to 6 GPa at the Institute for Solid State Physics, The University of Tokyo. We have taken the Fluorinert FC70 and FC77 (1:1 ratio) as PTM to generate high pressure, and the intramural pressure was calibrated from the structural transitions of Bi at room temperature [33]. The internal structure of the P-CAC consists of a constellation of six anvils (WC) around the cubic pyrophyllite gasket, which consists of the four external probe leads. The uniaxial pressure was produced by the six anvils, using a couple of guidance anvils at the top and bottom. The P-CAC was mounted in the double-layered glass dewar, which was then filled with liquid helium to attain the low temperature.

The HP-XRPD data at ambient temperature were recorded at the Xpress beamline [34] of the Elettra Synchrotron facility in Trieste, Italy. The pressure was generated using a membrane-driven symmetric diamond anvil cell (DAC) having a culet size of 400 μm (ϕ). For the membrane pressure control, a PACE-5000 automatic pressure controller was used. The sample chamber was prepared by intending a 200 μm thickness stainless steel gasket to 50 μm and by making a 160 μm through a hole in the center of the intended area by a spark-eroder. This gasket was placed in between the culet of both diamonds, and DAC was locked with sample and ruby chips (10 μm) together with PTM as a silicon oil. To determine the actual generated pressure within the DAC in-situ ruby fluorescence technique was employed. HP diffraction measurements were performed using a circular, monochromatic beam with a cross-sectional ϕ of 40 μm and a wavelength λ= 0.4957 Å. The PILATUS3 S 6M large-area detector was used to collect the diffracted X-ray from the sample, and Dioptas software [35] was used to convert the collected diffracted image pattern to 2 θ vs. intensity. The lattice parameter and unit cell volume under pressure were derived by the Rietveld refinement technique using the GSAS-II program [36].

## **Results and discussion: -**

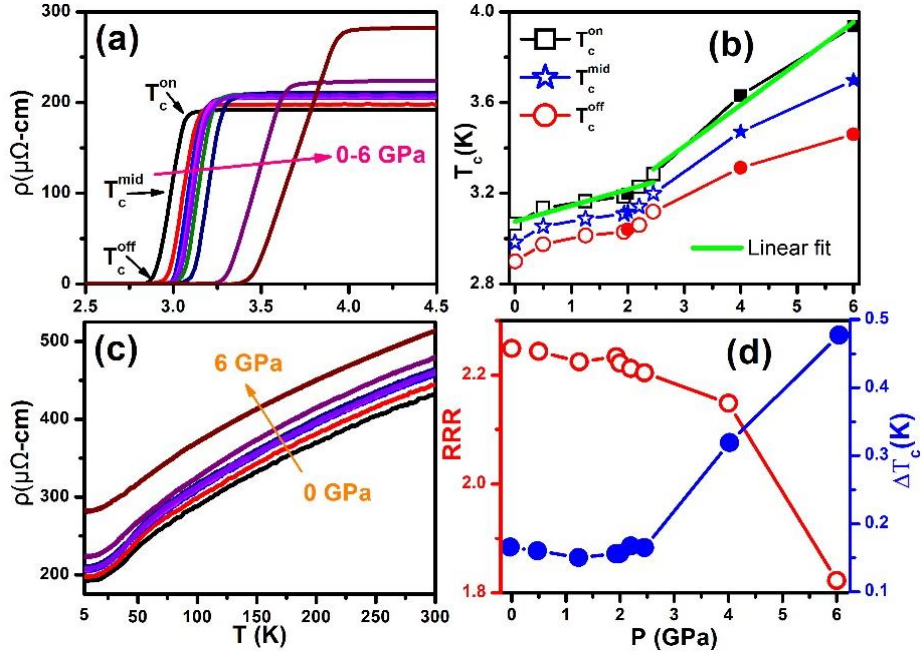

Figure. 1(a) *ρ(T)* in the superconducting transition region under various pressures (0-6 GPa) at zero magnetic fields, (b) Pressure dependence of $T_c^{on}, T_c^{mid}, \& T_c^{off}$, the open symbol represents piston-cylinder pressure cell data, and the solid symbol represents the P-CAC data, (c) Normal state *ρ(T)* curve from 0 to 6 GPa, (d) Pressure dependence of residual resistivity ratio and superconducting transition width of LaPtGe.

The *ρ(*T*)* around the superconducting transition region under various fixed pressures at zero magnetic fields is shown in Figure 1(a). The onset of superconducting transition temperature ($T_c^{on}$) is measured from the intersection of two lines drawn between the normal state and superconducting transition region, respectively. The $T_c^{on}$ was observed around 3.06 K at ambient pressure, which is similar to the previous report [22]. Also, a clear indication of enhancement in $T_c$ with broad superconducting transition was observed with increasing pressure. As mentioned in Figure 1(a), $T_c^{on}, T_c^{mid}, \& T_c^{off}$ are extracted and plotted for their corresponding pressure in Figure 1(b). The $T_c^{mid}$ is taken from the 50 % of superconducting transition and $T_c^{off}$ is taken from the intersection of two lines through the zero electrical resistivity state and superconducting transition region. Interestingly, it is found that the enhancement of $T_c^{on}$ is observed from 3.06 K (0 GPa) to 3.94 K for pressure 6 GPa with two distinct positive pressure coefficients *dT$_c$/dP*= 0.071 K/GPa and 0.183 K/GPa below and above ~3 GPa as shown in linear fit of Figure 1(b). However, the similar family compound CaPtAs (*dT$_c$/dP*= 0.071 K/GPa) [26] only exhibits the monotonic enhancement of $T_c$. Further, the monotonic change of $T_c$ is shown by a LaAlSi (*dT$_c$/dP*= 0.062 K/GPa) and LaAlGe (*dT$_c$/dP*= 0.03 K/GPa) [37] at very high pressure above 60 GPa. Besides, the careful observation of the pressure dependence of $T_c$ of LaPtGe shows a noticeable change in the enhancement of $T_c$ around 3 GPa.

Figure 1(c) shows a normal state *ρ(T)* under pressure from 0 to 6 GPa, and it indicates the enhancement of resistivity with pressure similar to the CaPtAs [26], and CePtSi [38]. The decrease of DOS could be the possible reason for the enhancement of $T_c$ with external pressure [11] [39]. In addition, the residual resistivity ratio (RRR) is calculated from $\rho^{300\,K}/\rho^{5\,K}$ and it is plotted with its corresponding pressures up to 6 GPa is shown in Figure 1(d). The observed value of RRR at both ambient and high pressure is comparable with other non-magnetic NCS superconductors like $Re_{24}Ti_5$ [40], $Nb_{0.18}Re_{0.82}$ [41,42], $Re_6Zr$ [43], $TaRh_2B_2$ [44,45], $NbRh_2B_2$ [44], $Re_6Hf$ [46], $Re_{5.5}Ta$ [47]. Further, the pressure dependence of superconducting width was calculated using $\Delta T_c = T_c^{on} - T_c^{off}$ and it is plotted in Figure 1(d). Both RRR and $\Delta T_c$ were almost unchanged, with pressure up to ~3 GPa. But around 3 GPa, an apparent change was observed in the trend of both pressure-dependent RRR and $\Delta T_c$, which is comparable with the enhancement of pressure-dependent $T_c$ up to 6 GPa.

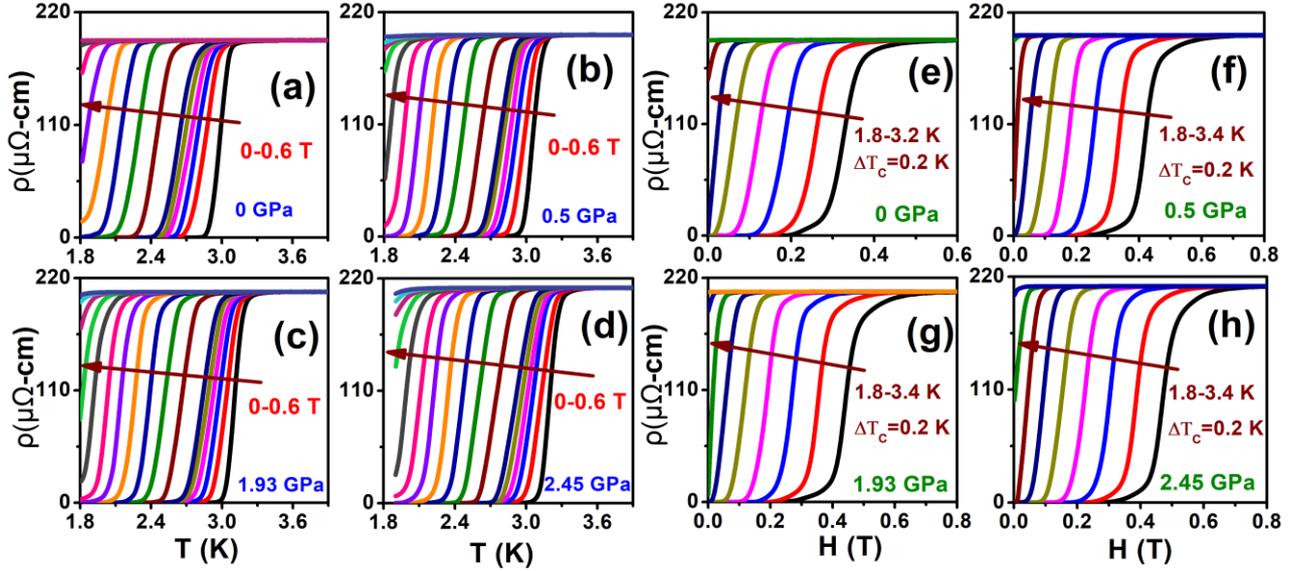

Figure. 2(a-d) Temperature dependence of resistivity in the superconducting transition region with various external magnetic fields and (e-h) Field dependence of resistivity around superconducting transition zone at various fixed temperature points under pressure 0, 0.5, 1.93, and 2.45 GPa respectively.

To determine the $H_{c2}(0)$ and the role of the magnetic field in the $T_c$ under pressure, we measured the $\rho(T)$ with various magnetic fields and $\rho(H)$ at different temperatures for various fixed pressures up to 2.45 GPa. The $\rho(T)$ from the temperature range 1.8 K to 3.9 K under various magnetic fields (0 to 0.6 T) and $\rho(H)$ from 0 to 0.8 T under various fixed temperatures from 1.8 K to 3.2 K are shown in Figure 2(a-d) and Figure 2(e-h), respectively.

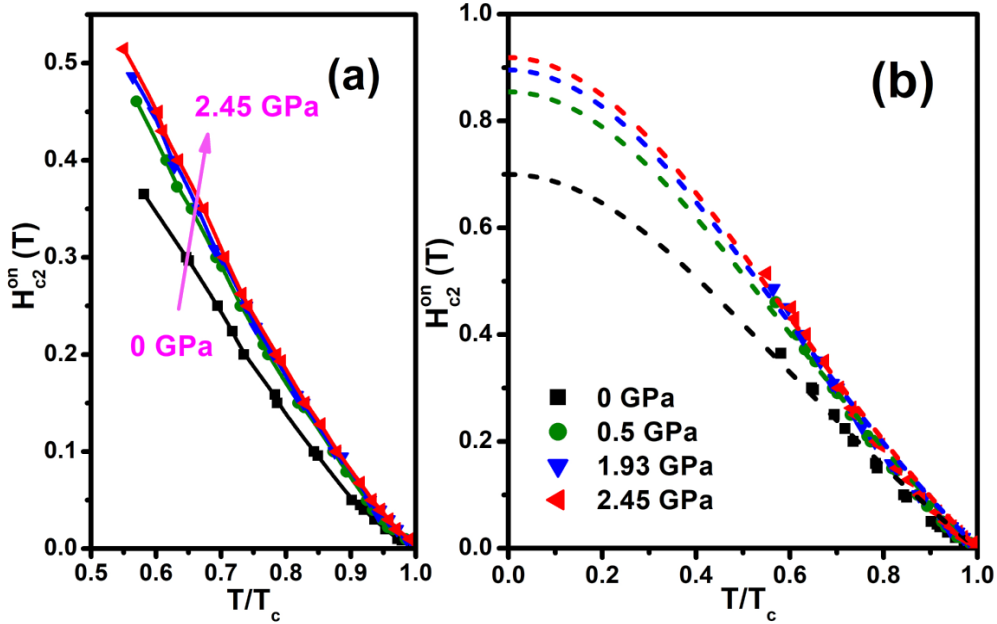

Figure. 3 (a & b) Normalized temperature dependence of upper critical field, $H_{c2}(T)$, from $\rho(T)$ & $\rho(H)$ under different pressure from 0 to 2.45 GPa. The dotted lines represent the GL fit using the relation, $H_{c2}(t) = H_{c2}^{GL}(0)[(1-t^2)/(1+t^2)]$.

For each pressure, the temperature dependence of the upper critical field ($H_{c2}(T)$) was extracted from $\rho(T)$ & $\rho(H)$ and plotted with normalized temperature ($t=T/T_c$) as shown in Figure 3(a). It clearly reveals the enhancement of $H_{c2}(0)$ with pressure. The careful analysis of $H_{c2}(T)$ shows the upward augmentation of $H_{c2}(T)$ by increasing pressure from 0 GPa and 2.45 GPa, as shown in Figure 3(a). To understand the precise evaluation of the upper critical field with pressure, for each pressure, we have calculated an $H_{c2}(0)$ by fitting the $H_{c2}(T)$ with the normalized temperature using the following Ginzburg-Landau (GL) relation, $H_{c2}(T) = H_{c2}^{GL}(0)[(1-t^2)/(1+t^2)]$ [48]. The $H_{c2}^{GL}(0)$ is determined by using GL fitting (dotted line) for ambient and HP up to 2.45 GPa as shown in Figure 3(b). In addition, the orbital upper critical field $H_{c2}^{orb}(0)$ is calculated using the Werthamer-Helfand-Hohenberg (WHH) relation, $H_{c2}^{orb}(0) = -0.693\, T_c (dH_{c2}/dT)_{T_c}$ [49]. Where, $\left(\frac{dH_{c2}}{dT}\right)_{T_c}$ is the slope obtained from the $H_{c2}(T)$ curve, where $T = T_c$.

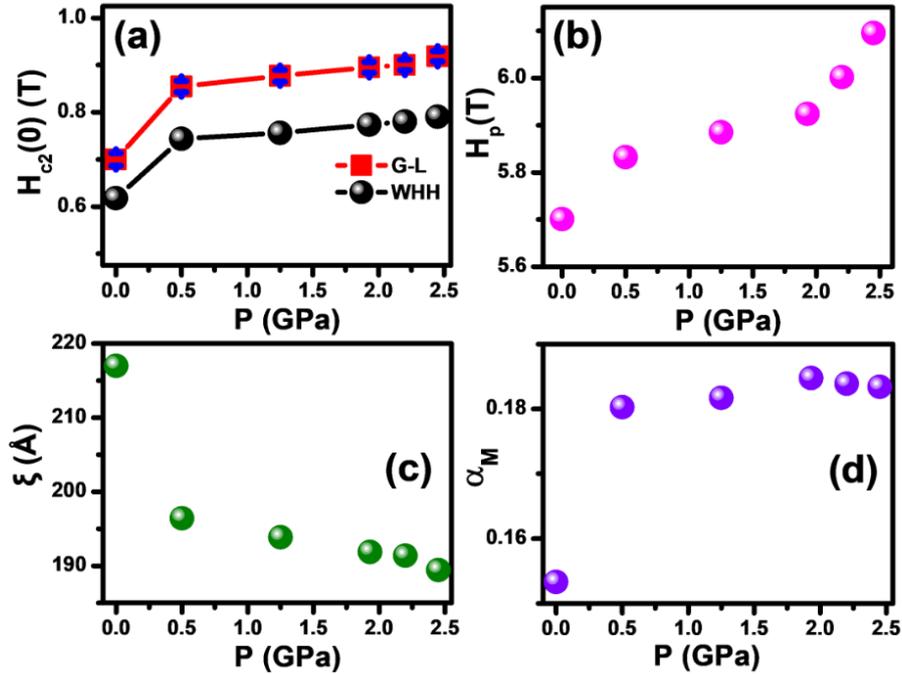

Figure. 4(a) Pressure dependence of calculated upper critical field at absolute zero using *GL* and *WHH* equation, (b) Pressure dependence of Pauli limiting field, (c) Pressure dependence of coherence length, (d) Pressure dependence of Maki parameter.

The calculated upper-critical fields [$H_{c2}^{GL}(0)$ & $H_{c2}^{orb}(0)$] at absolute zero up to 2.45 GPa are shown in Figure 4(a), which clearly reveals an increase of $H_{c2}(0)$ with pressure. The enhancement of $H_{c2}^{GL}(0)$ from 0.7 T to 0.92 T was observed by increasing pressure from 0 GPa to 2.45 GPa. Also, the Pauli upper critical field, $H_p$, was calculated using the relation, $H_p=1.86\times T_c$ [50] for all pressures up to 2.45 GPa. The pressure dependence of $H_p$ clearly indicates $H_{c2}(0) < H_p$, and it confirms the s-wave superconductivity of LaPtGe. Further, the GL superconducting coherence length ($\xi_0$) was calculated for all pressures using the following relation, $H_{c2}^{GL}(0) = \Phi_0/2\pi\xi_0^2$ (Where, $\Phi_0 = 2.07\times 10^{-15}$ Wb is the flux quantum). The pressure dependence of $\xi_0$ up to 2.45 GPa was plotted in Figure 4(c) and it indicates the decrease of $\xi_0$ with pressure. In addition, the Maki parameter was calculated using the relation, $\alpha_M = \sqrt{2}H_{c2}^{orb}(0)/H_p(0)$ [51] at both ambient and HP. The increase of $\alpha_M$ up to 2.45 GPa than 0 GPa is shown in Figure 4(d) and the small value of $\alpha_M$ less than one indicates the BCS superconductivity. However, the careful analysis of pressure-dependent $H_{c2}(0)$ shows interesting results in the initial enhancement of $H_{c2}(0)$ from 0 to 0.5 GPa, followed by a moderate increment observed from 0.5 GPa to 2.45 GPa under pressure. This pressure-dependent tendency of $H_{c2}(0)$ can be correlated with the tendency of pressure-dependent $T_c$, even in the 0 to 0.5 GPa trend and 2.2 to 2.45 GPa. Vice versa the noticeable enhancement of $H_{c2}(0)$ is expected above 2.45 GPa similar to the enhancement of $T_c$.

Further, the noticeable decrement in RRR and increment in $T_c$, $\Delta T_c$ around 3 GPa, strongly suggests the significant role of pressure in the enhancement of $H_{c2}(0)$ and/or the upward augmentation of $H_{c2}(T)$ above 2.45 GPa, and it was expected as multi gap feature like LaPtSi [23]. Moreover, the sensible growth in the upward tendency of $H_{c2}(T)$ with pressure might be raised by the lattice contraction. To shed light on the structural aspects, we performed the HP-XRPD studies.

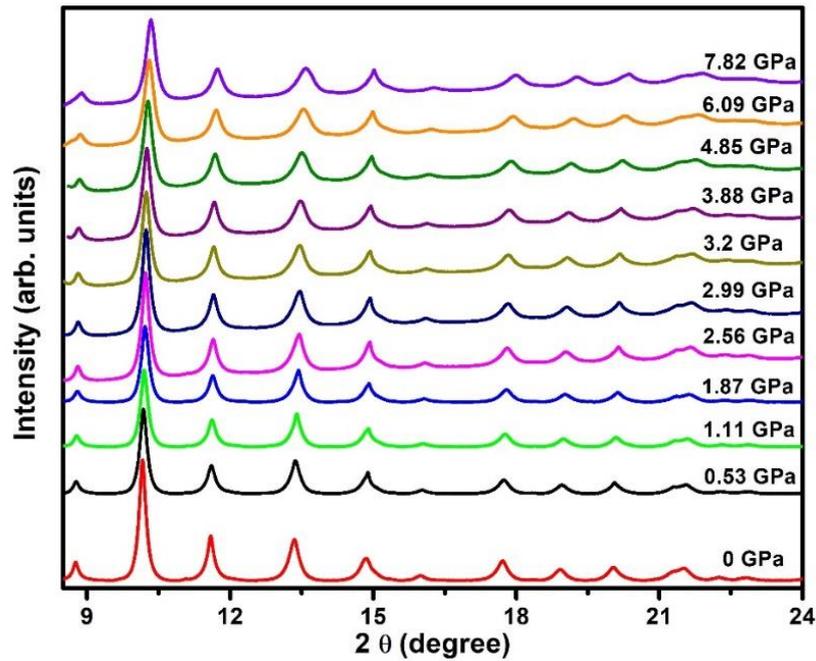

Figure 5. Measured HP-XRPD patterns of LaPtGe up to 7.82 GPa. The patterns are stacked vertically for a better presentation.

Figure 5 shows the observed XRPD patterns for different pressures up to 7.82 GPa using silicon oil as a PTM. For better comparison, the patterns were stacked vertically. It shows a regular shift to a higher 2θ with increasing pressure in accordance with a regular shrinkage of the lattice. The data shown in Figure 5 does not evidence any structural phase transitions up to the highest pressure measured (7.82 GPa).

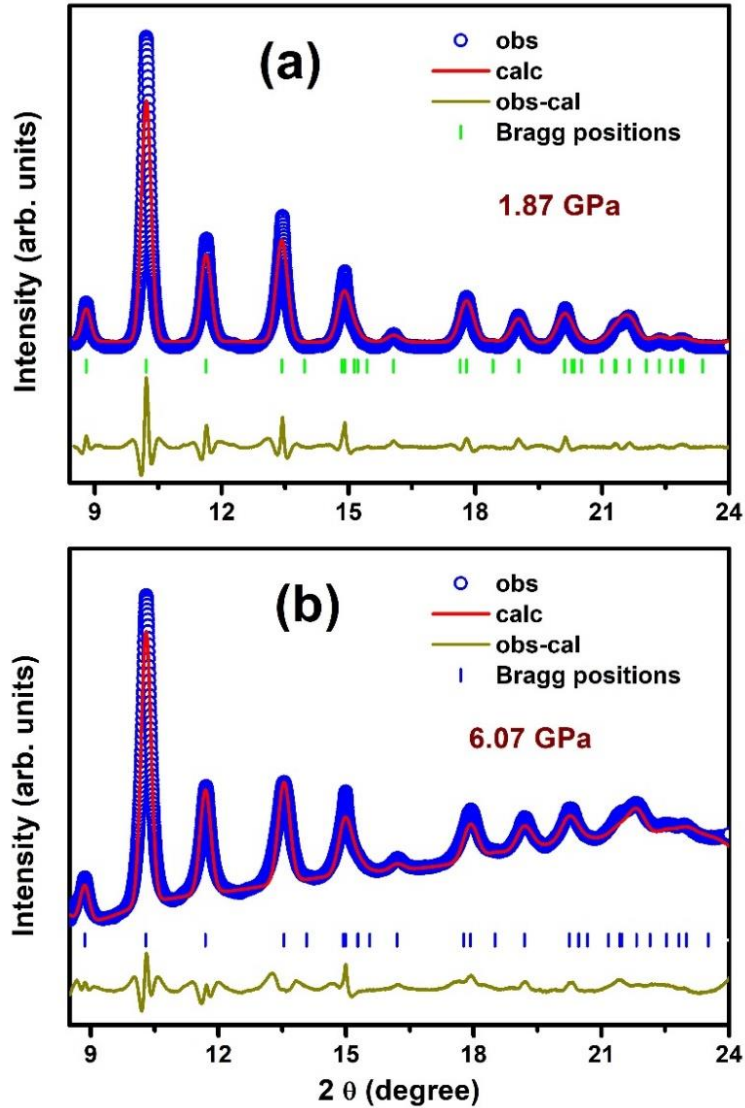

Figure 6. Measured HP-XRPD of LaPtGe at (a) 1.87 and (b) 6.07 GPa pressure with synchrotron radiation using λ=0.4957 Å (symbols) together with the Rietveld refinement results (solid red lines). The Bragg peak positions are shown in vertical green lines, and the dark yellow line shows the difference between observation and calculations.

Figure S1 (in SI) shows the measured XRPD pattern of LaPtGe at ambient pressure using synchrotron radiation (wavelength λ=0.4957 Å) together with the results of the Rietveld refinement analysis. Figure 6 shows the measured XRPD pattern of LaPtGe at 1.87 and 6.07 GPa. In both figures, the blue circle symbols, red solid line, and green vertical bars represent the experimental data, refinement result, and Bragg peak positions of the parent phase, respectively. At ambient conditions, all the Bragg-peaks can be indexed to a tetragonal structure with space group *I4₁md* (No.109) (see Figure S1) and lattice parameters $a = b = 4.265$ Å, $c = 15.02$ Å ($V = 273.279$ Å$^3$) respectively, in good agreement with the earlier report [22].

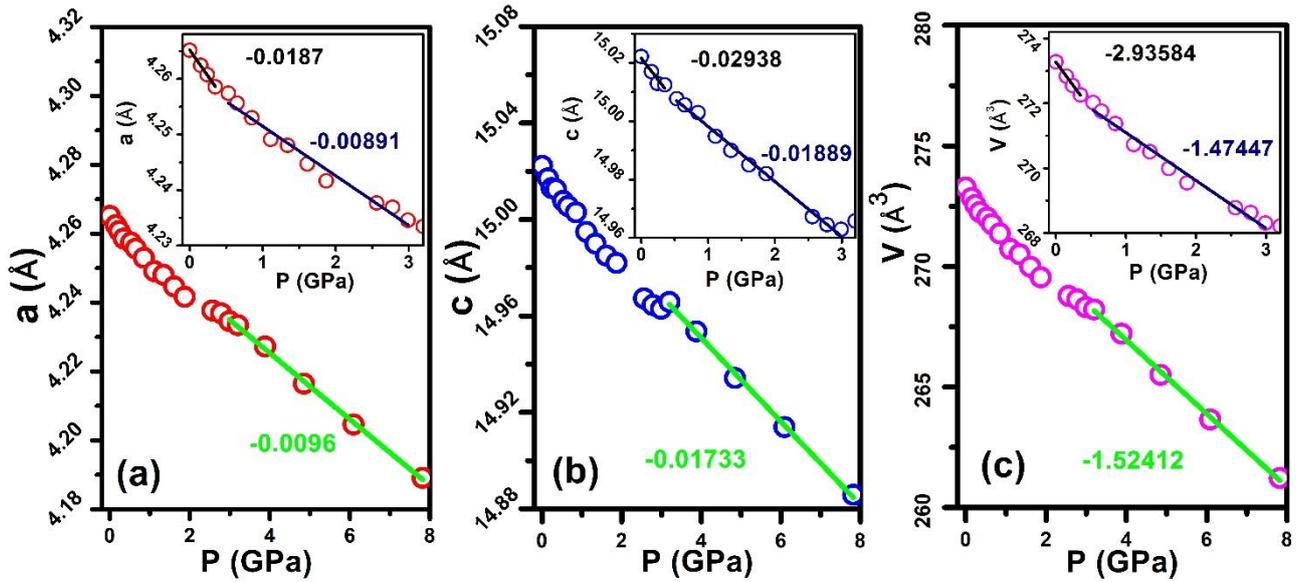

Figure 7. Pressure dependence of the lattice parameters (a) $a = b$ (Å), (b) $c$ (Å), and (c) $V$ (Å$^3$). The insets of each figure show the lattice parameters in the low-pressure region (below 3 GPa). The symbols are the experimental data, the solid lines are the linear fits, and the numbers close to the lines are the slopes (in Å/GPa).

Figure 7 shows the pressure dependence of lattice parameters $a=b$ (Å) and $c$ (Å) of the LaPtGe system. More interestingly, in both cases, data above ~3 GPa are found to clearly fall in a line, in contrast to the low-pressure points (inset of Figure 7). For the high-pressure side (> 3 GPa), the rates for the c-axis and a-axis were respectively -0.0173 and -0.0096 Å/GPa (see Figure 7 main panel). The data below 3 GPa does not fit well with the single linear fit, but it follows a two-linear dependence between 0 to 0.5 GPa and 0.5 to 3 GPa. The rate of shrinkage of the *c*-axis is almost double that of the *a*-axis, indicating a strong anisotropy in the in-plane and out-of-plane compression. However, careful observation of the lattice compression above 3 GPa shows anisotropic shrinkage of the *a*-axis and *c*-axis, such as ~108 % and ~92 %, than its compression between 0.5 GPa to 3 GPa. Furthermore, the keen observation of the lattice compression in the different regions, such as 0 to 0.5 GPa, 0.5 to 3 GPa, and above 3 GPa, evidently supports the enhancement of $T_c$ in the corresponding regions. This clearly supports the enhancement of $T_c$ and inducible $H_{c2}(0)$ above 2.45 GPa.

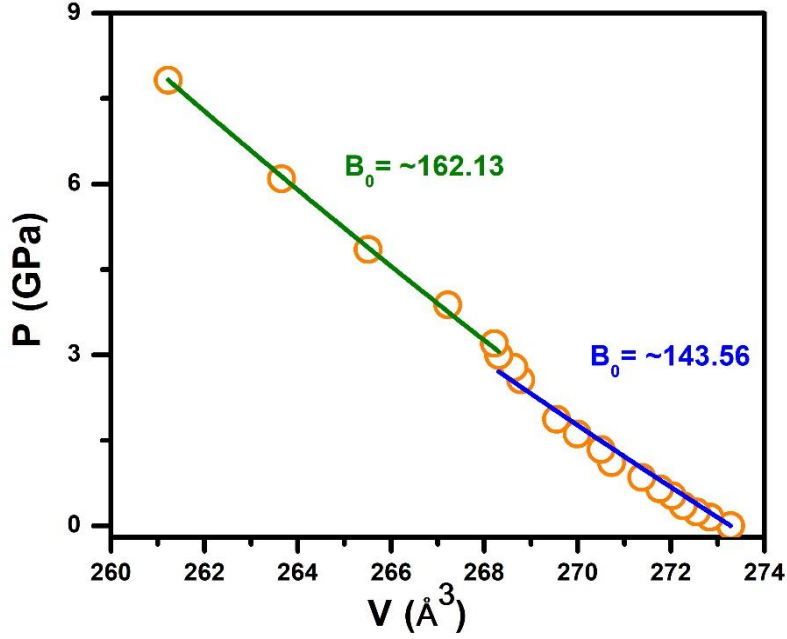

Figure 8. Pressure dependence of the unit-cell volume $V$ (Å$^3$), the green and blue solid lines, respectively, represent a second-order Birch-Murnaghan equation of state fit to the data below and above ~3 GPa. The numbers close to the lines are the bulk-modulus (in GPa) obtained from the fit.

The pressure dependence of unit-cell volume $V$ is presented in Figure 8. With increasing pressure, $V$ does not follow a smooth continuous curve. It shows two distinct pressure regimes similar to the lattice parameters (Figure 7). This motivated us to treat the data as two discrete regimes, following the similar pressure regimes used for the lattice parameters: ~3 GPa and above > 3 GPa. To calculate $B_0$, the pressure dependence of unit cell volume was fitted (below and above ~3 GPa) using the Birch-Murnaghan (BM) equation of state, $P(V) = \frac{3B_0}{2}\left[\left(\frac{V_0}{V}\right)^{\frac{7}{3}} - \left(\frac{V_0}{V}\right)^{\frac{5}{3}}\right]\left\{1 + \frac{3}{4}\left(B_0' - 4\right)\left[\left(\frac{V_0}{V}\right)^{\frac{2}{3}} - 1\right]\right\}$. In Figure 8, the solid green line and the dash-dotted red line represent, respectively, the fitting of the second-order BM equation in the high and low-pressure regimes. This analysis provides the two different $B_0$, such as ~144 GPa and ~162 GPa, for the low- and high-pressure regimes, respectively. Compared to reported values of $B_0$ for similar systems, this is slightly higher; for example, 108 GPa is reported for CePtSi [38] and 106 GPa is reported for LaAlSi [37]. We also note that the calculated value of 188.3 GPa at 5 GPa for the ThIrGe [52].

Interestingly, analyzing the HP-XRPD data reveals a clear distinction between the pressure dependence of lattice values above and below ~3 GPa. This distinct subtle lattice change is correlated with the trend change observed around 3 GPa in the pressure dependence of $T_c$, RRR, and $\Delta T_c$ from resistivity. The compression of unit cell volume may increase the electron-phonon coupling strength, and that might be a possible reason for the enhancement of $T_c$ [11] [39] [53]. So, the robust behaviour observed from HP-XRPD around 3 GPa could be related to the resistivity. Noticeably, this upward augmentation of $H_{c2}(T)$ with pressure (Figure. 3a) is comparable with the similar family compound LaPtSi [23]; the LaPtSi shows a clear upward trend at ambient pressure. However, this upward trend in LaPtGe is not as clear as LaPtSi, but it is sensible by increasing pressure. Even though, the band structure analysis suggests a comparably small value of DOS at Fermi level ($N(E_F)$) for LaPtGe than the LaPtSi [24]. Comparably larger unit cell volume of the LaPtGe [22] [24] than the LaPtSi [24] might be a possible reason for this difference. So, the enhancement of $T_c$ and upward augmentation of $H_{c2}(T)$ might be due to the lattice compression and the decrease of DOS with pressure. It is comparable with the multi-gap behaviour in the parent compound LaPtSi [23].

## **Conclusion: -**

In conclusion, the pressure effect on the superconducting, electronic and structural properties of the NCS LaPtGe compound was studied through electrical resistivity measurements up to 6 GPa and XRPD up to 7.8 GPa. The enhancement of $T_c$ from 3.06 K to 3.94 K was observed under pressure with two distinct trends, below and above ~3 GPa, respectively. The slight increase in $T_c$ with pressure up to 3 GPa, at a rate of 0.071 K/GPa. Above this pressure, however, the rate increased by about 2.5 times to 0.183 K/GPa. Further, the HP-XRPD experiments using the synchrotron show the two discrete trends in the pressure dependence of lattice parameters below and above ~3 GPa without a structural phase transition up to 7.82 GPa. And BM equation of state fit shows two different bulk modulus, ~144 GPa (P < 3 GPa) and ~162 GPa (P > 3 GPa). In addition, the magnetotransport measurement under pressure up to 2.45 GPa demonstrates the increase in the upper critical field ($H_{c2}(0)$) from 0.7 T (0 GPa) to 0.92 T (2.45 GPa), and it can be correlated well with the pressure dependence of $T_c$. The calculated $H_{c2}(0)$ does not exceed the Pauli limit at both ambient and high pressure, which suggests the *s*-wave superconductivity remains with pressure.

Interestingly, the noticeable upward augmentation of $H_{c2}(T)$ up to 2.45 GPa does not entirely rule out the possibility of multi-gap superconductivity in LaPtGe at high-pressure, identical structured LaPtSi. The distinct enhancement of $T_c$ around 3 GPa, the pressure dependence of RRR, and the noticeable shift in lattice parameters around 3 GPa strengthen the possibility of apparent upward augmentation of $H_{c2}(T)$ and multi-gap nature above 2.45 GPa. This change in $T_c$ can be correlated with the enhancement of electron-phonon interaction and the change of DOS at the Fermi level. Also the monotonic shift of pressure-dependent activation energy $U_0$ up to 2.45 GPa determined by the Arrhenius equation states that the change in $T_c$ is not due to grain boundaries. Furthermore, the TAFF analysis suggests the coexistence of a single and collective vortex state at ambient and high pressure (Refer to supplementary material). Also, it indicates the dominance of a single vortex state over the collective vortex states with pressure up to 2.45 GPa, and it can be correlated with the enhancement of $T_c$ with pressure. However, high-pressure low-temperature measurements (below 1.8 K and above 6 GPa) in LaPtGe are necessary to conclude the pressure-dependent of $H_{c2}(0)$.

**Acknowledgements: -**

Author S.M. thanks SERB for the financial support through OVDF (ODF/2018/000184) and acknowledges ISSP, The University of Tokyo, Japan, for providing the experimental facilities. Further, the author S.M. thanks ICTP for the travel support and high-pressure diffraction experiments at the Xpress beamline of the Elettra Sincrotrone Trieste (proposal number **20230257**) and RUSA 2.0 for the fellowship. Author A.S. acknowledges Indus-2 beamline (RRCAT), SERB, DAE-BRNS, and TANSCHE.